\documentclass[aps,pra,reprint,nofootinbib,superscriptaddress,twocolumn,showpacs,showkeys,longbibliography,amsmath,amssymb]{revtex4-1}
\usepackage{graphicx}
\usepackage{dcolumn}
\usepackage{bm}
\usepackage{braket}
\usepackage{subfigure}
\usepackage[colorlinks,bookmarks=false,citecolor=blue,linkcolor=red,urlcolor=blue]{hyperref}
\usepackage[english]{babel}

\begin{document}
\preprint{APS/123-QED}

\title{Dynamics of two dark solitons in a polariton condensate under non-resonant pumping}

\author{Yiling Zhang}
\affiliation{Department of Physics, Zhejiang Normal University, Jinhua 321004, People's Republic of China}
\author{Chunyu Jia}
\affiliation{Department of Physics, Zhejiang Normal University, Jinhua 321004, People's Republic of China}
\author{Zhaoxin Liang}\email[Corresponding author:~] {zhxliang@zjnu.edu.cn}
\affiliation{Department of Physics, Zhejiang Normal University, Jinhua 321004, People's Republic of China}

\date{\today}
\begin{abstract}
We theoretically investigate the dynamics of two dark solitons in a polariton condensate under nonresonant pumping, based on driven dissipative
Gross-Pitaevskii equations coupled to the rate equation.  The equation of motion of the relative center position of two-dark soliton is obtained analytically by using the Lagrangian approach. In particular,  the analytical expression of the 
effective potential between two dark solitons is given. The resulting equation of motion captures how the open-dissipative character of a
polariton Bose–Einstein condensate affects the properties of dynamics of two-dark soliton, i.e., two-dark soliton relax
by blending with the background at a finite time. 
We further simulate the relative motion of two dark solitons numerically with the emphasis on how two solitons' motion being manipulated the initial velocity, which are in excellent agreement with the analytical results. The prediction of this work is sufficient for the experimental observations within current facilities.
\end{abstract}

\maketitle


{\it Introduction.---} At present, there are significant research interests in exciton-polariton Bose-Einstein condensates (BEC) in semiconductor microcavities as a novel platform for realization and investigation of nonlinear physics~\cite{exciton1,exciton2,exciton3,exciton4}. On the one hand, at the mean-field level, the static and dynamic properties of a polarition BEC can be well described by the nonlinear Schr\"odinger equation or Gross-Piteavskii equation (GPE), which has been a paradigm of theoretical and experimental studies of coherent nonlinear dynamics~\cite{coherent1, coherent2, coherent3, coherent4}.  On the other hand, a polariton condensate is intrinsically non-equilibrium, with coherent and dissipative dynamics occurring on an equal footing. This has provided a new stage for practical applications of the GPE. Up to now, the non-equilibrium nature of the polariton has resulted in numerous intriguing nonlinear phenomena in a polariton condensates~\cite{dissipative1,dissipative2}. 

In the quest for novel scenarios that display combined effects of dissipation and nonlinearity on the nonlinear phenomena, the study of soliton in polariton condensates is among the hottest topics, with an emphasis on capturing the non-equilibrium nature of the soliton with no analog in the static counterpart~\cite{PoDark1,PoDark2,PoDark3,PoDark4,PoDark5,PoDark6,PoDark7,PoDark8,PoDark9,PoDark10}.  Generally speaking, soliton is a kind of self-reinforcing solitary wave, which is formed by the cancellation of nonlinear effect and dispersion effect in medium and it can maintain its shape during propagation~\cite{soliton0, soliton1, soliton2, soliton3}. The dark or bright soliton can exist provided the interaction is repulsive or attractive~\cite{schro1, schro2}. In a polariton condensate, the nonlinearity of the polariton condensate arises from the strongly and repulsively interacting excitons, where the interaction can be controlled via Feshbach resonance~\cite{NP2014,Feshbach2017}. A series of experiments have demonstrated the existence of the oblique dark solitons and vortices~\cite{vortex1,vortex2,vortex3}, or bright spatial and temporal solitons~\cite{sky1}. For example, in condensates created spontaneously under incoherently
pumping, the formation and properties of dark solitons have been investigated theoretically in Ref.~\cite{creation1}, and the existence of stable dark soliton trains has been reported in the non-resonantly driven spinor polariton BEC at one dimension~\cite{stable1}.  However, to our best knowledge, previous studies on solitons in a polariton condensate have been limited to the one-soliton problem, whereas the two-soliton problem in non-equilibrium polariton BEC has remained so far unexplored. It is the purpose of the present work to investigate how the interplay of nonlinearity, dispersion and dissipation affects the existence and properties of  two-dark soliton in a polariton BEC. 

In this Letter, we present the first analytical result on the two-dark soliton problem in the context of a polariton BEC formed under non-resonant pumping by solving the dissipative GPE. First, we use the variational approach and analytically derive the time evolution equations for the soliton parameters, i.e., the relative distance between solitons. We compare this analytical result with the numerical solutions for the trajectory of two solitons directly obtained from the dissipative GPE, finding a remarkable agreement between the two. Our results open a new route to observe stable two-solitons in non-equilibrium polariton BEC within current experimental facilities.

{\it Model.---} 
 \begin{figure*}
\centering{\Large}
\includegraphics[width=0.8\textwidth]{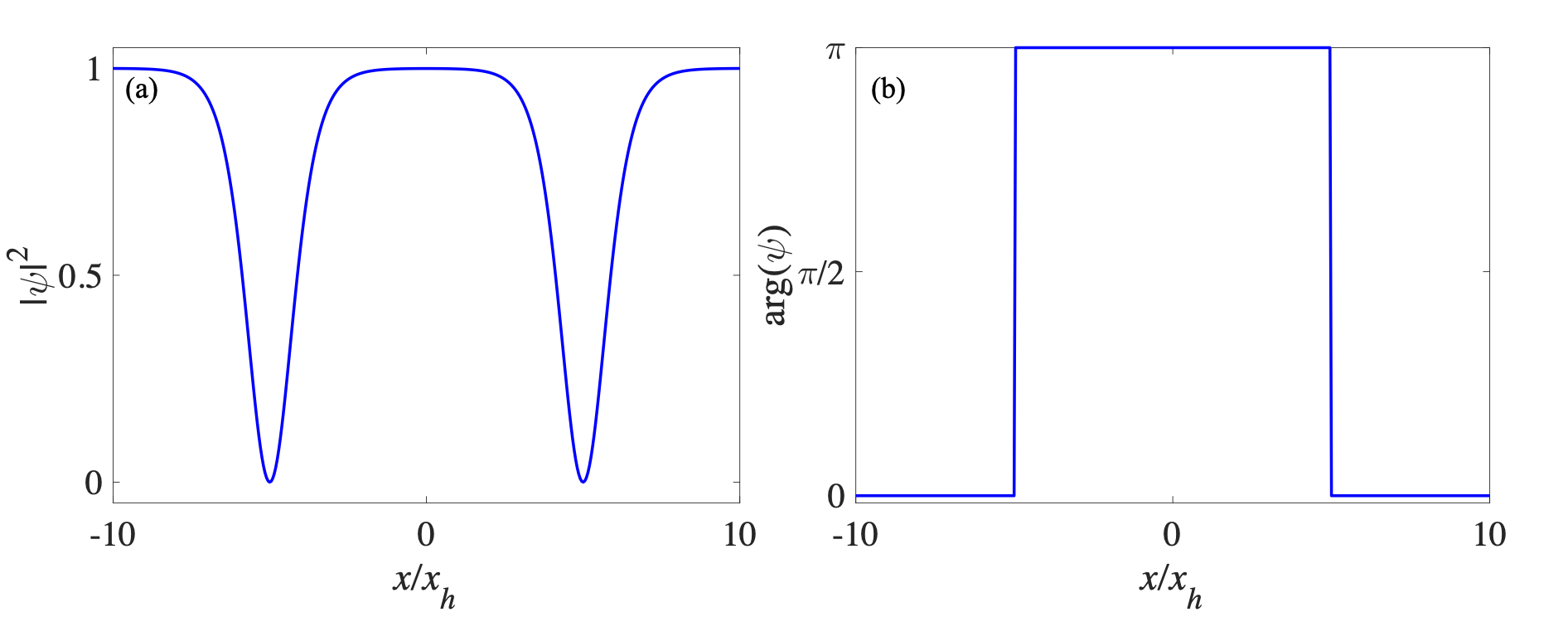}
\caption{(a) Schematics of the condensate density of $\left|\psi\right|^{2}$ and (b) phase of the condensate order parameter of $\arg(\psi)$ corresponding to a stationary one-dimensional two-dark-soliton solution in Eq. (\ref{Tdark}) with the
parameter of $v_s=0$. }
\label{a1}
\end{figure*}
We are interested in a exciton-polariton BEC under nonresonant pumping created in a wire-shaped microcavity similar to the one implemented in Ref. \cite{Wertz2010}  that bounds the polaritons to a quasi-one-dimensional (1D) channel. Within the framework of the mean field theory, the polariton field described by the condensate order parameter of $\psi(x,t)$ evolves along an effectively 1D  driven-dissipative GPE coupled to a rate equation for the density $n_{R}(x,t)$ of reservoir polaritons as follows \cite{dark_polariton1},
 \begin{eqnarray}
\!\!\!\!i\hbar\frac{\partial\psi}{\partial t}\!&=&\!\Big[\!-\!\!\frac{\hbar^{2}}{2m}\!\frac{\partial^2}{\partial x^2}\!+\!g_{C}|\psi|^{2}\!+\!g_{R}n_{R}\!+\!\frac{i \hbar}{2}(Rn_{R}\!-\! \gamma_{C})\Big]\psi,\label{psi}\\ 
\frac{\partial n_{R}}{\partial t}&=&P-\left(\gamma_{R}+R|\psi|^{2}\right)n_{R}. \label{rate}
 \end{eqnarray}
 In Eqs. (\ref{psi}) and (\ref{rate}), the $m$ is the effective mass of lower polaritons and  $P$ is the rate of an off-resonant continuous-wave pumping; $\gamma_C$ and $\gamma_R$ describe the lifetime of the condensate and reservoir polaritons respectively, and $R$ is the stimulated scattering rate of reservoir polaritons into the condensate;  $g_{C}$ and $g_{R}$ characterize the strength of nonlinear interaction of polaritons and the interaction strength between reservoir and polariton respectively. Note that the parameters of $g_{C}$, $g_{R}$, and $R$ have been rescaled into the one-dimensional case by the width $d$ of the nanowire thickness  as that $g_{C}\rightarrow g_{C}/\sqrt{2\pi d}$, $g_{R}\rightarrow g_{R}/\sqrt{2\pi d}$, $R\rightarrow R/\sqrt{2\pi d}$).
 
The emphasis and value of this work are to take account of the intrinsic dissipation and look for the possibility of the existence and dynamics of two-dark soliton in a polariton condensate. 
 It's well known that the dark soliton is characterized with a localized dip in the condensate density with an associated
phase gradient. Hence, we first need to determined the steady state of the model system based on Eqs. (\ref{psi}) and (\ref{rate}), which serves as the density background of the dark soliton's propagation. 
Directly following Ref. \cite{dark_polariton1},  the stable condensate appear under the condition of the pumping $P$ in Eq. (\ref{rate}) being bigger than a critical value of $P_{\text{th}}=\gamma_R\gamma_C/R$. In such, 
the steady homogeneous condensate and reservoir densities are expressed as follows: $n^0_C=(P-P_{\text{th}})/\gamma_C$ and $m^0_R=\gamma_C/R$.
 
For convenience, we proceed to rescale Eqs.~(\ref{psi}) and (\ref{rate}) into the dimensionless form.  In more details, we rescale $\psi\rightarrow \psi/\sqrt{n^0_C}$ and introduce $m_R=n_R-n^0_R$; as a result, Eqs.~(\ref{psi}) and (\ref{rate}) can be rewritten as
 \begin{eqnarray}
i\frac{\partial\psi}{\partial t}&+&\frac{1}{2}\frac{\partial^2 \psi}{\partial x^2}\!-\!\left(|\psi|^{2}\!-\!1\right)\psi=(\bar{g}_{R}m_{R}\!+\!\frac{i}{2}\bar{R}m_{R})\psi,\label{Spsi}\\
\frac{\partial m_{R}}{\partial t}&=&\bar{\gamma}_{C}\left(1-|\psi|^{2}\right)-\bar{\gamma}_R m_{R}-\bar{R}|\psi|^{2}m_{R}. \label{Srate}
\end{eqnarray}
Here, the dimensionless parameters are given by $\bar{g}_{R}=g_{R}/g_{C}$, $\bar{\gamma}_{C}=\gamma_{C}\bar{\gamma}_{R}/\gamma_{R}$ and $\bar{R}=\hbar R/g_{C}n^0_C$. Moreover, the time $t$ and the space coordinate $x$ are measured in the units of $\tau_0=\hbar g n^0_C$ and $x_h=\sqrt{\hbar^2/m g n^0_C}$.  Note that Equations (\ref{Spsi}) and (\ref{Srate}) are the starting point of investigating the two-dark soliton problem in the context of the polariton BEC.
The non-equilibrium nature of model system is captured by the parameters of $\bar{R}$ in the right hand of Eq. (\ref{Spsi}). Blow, we are theoretically interested in how the nonequilibrium nature affecting the dynamics of two-dark soliton.

{\it Two-dark Soliton---} Before investigating the effects of the non-equilibrium nature of model system characterized by $\bar{R}$ on the two-dark soliton solution, we first briefly review some important features of the normal GPE, corresponding to Eq. (\ref{Spsi}) with $\bar{g}_R=\bar{R}=0$. Under the boundary condition of $\psi\rightarrow 1$ as $|x|\rightarrow \infty$, the two-dark-soliton solution can be written as~\cite{Lagrangian2}
\begin{equation}
\psi=\left(B\tanh x_{+}-iA\right)\left(B\tanh x_{-}+iA\right),\label{Tdark}
 \end{equation}
with $A^2+B^2=1$. Here $x_{\pm}=B\left(x\pm x_{0}\right)$ is referred as to the center position of two dark solitions and $2x_0$ can be treated as the relative center position between two solitons. In general, we can obtain that $x_0=v_s t$, $A=v_s$ and $B=\sqrt{1-v^2_s}$ with $v_s$ being the velocity of the dark soliton. In order to better understand the trial wavefunction of two dark solitons in Eq. (\ref{Tdark}), we have plotted the density profile with the parameter of $v_s=0$ in Fig. \ref{a1} (a) characterized that the phase of the two dark soliton solution $\psi(x,t)$ has $\pi $ phase jump profile (see Fig. \ref{a1} (b)). 
 
 \begin{figure*}
\centering{\Large}
\includegraphics[width=0.8\textwidth]{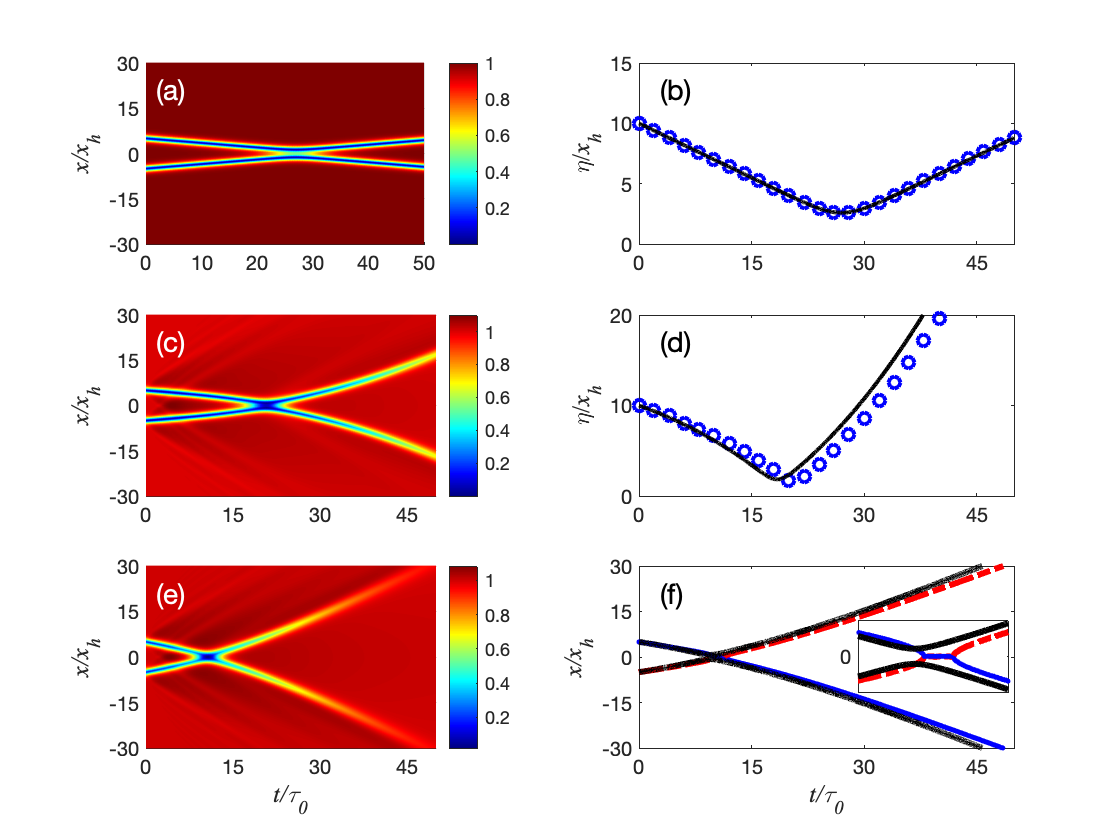}
\caption{Dynamics of 1D two-dark soliton in a polariton BEC with the different initial velocity. Left panel: the contour plots of the two dark solitons of  $\left|\psi\right|^{2}$ are shown; Right panel: the equations of motion of relative center mass of two dark soliton are plotted corresponding to the analytical results (solid curves) in Eq. (\ref{EOM}) and the numerical results (dotted curved)  by solving equation (\ref{Spsi})-(\ref{Srate}).The solid black lines are calculated using the analytical results of the relative motion of two-dark soliton $\eta =2x_0$ in Eq. (\ref{EOM}) in (b), (d) and (f). For the parameters: (a-b) $v_{s}=0.15$, $\bar{g}_{R}=\bar{\gamma}_{C}=\bar{\gamma}_{R}=\bar{R}=0$. In other plots, we have chosen $\bar{g}_{R}=2$, $\bar{\gamma}_{C}=3$, $\bar{\gamma}_{R}=15$ and $\bar{R}=1.5$, for (c-d) $v_s=0.15$; (e-f) $v_s=0.4$.}\label{Fig2}
\end{figure*}

Adding the open-dissipative effects as captured by $\bar{R}$ introduces an external perturbation of the two-dark-soliton in Eq. (\ref{Tdark}), which leads to two immediate consequences: first, all the parameters of two-dark-soliton solution in Eq. (\ref{Tdark}) become slow functions of time, i. e. $A\rightarrow A(t)$, $B\rightarrow B(t)$, and $x_0\rightarrow x_0(t)$,  while the functional form of the soliton remains unchanged, which is at heart of the Lagange approach of quantum dynamics of two-dark soliton in the presence of perturbation; second, it's supposed that the two-dark-soliton will relax by blending with the background at a finite time. In such, we will rely on the Lagarange approach of the dark soliton perturbation theory which allows for the analytical treatment of the effects of the right hand in Eq. (\ref{Spsi}).

We focus on the relative center mass position of two-dark-soliton solution corresponding to  the time variation of the parameter of $x_0(t)$. As such, we can obtain the equation of motion of $x_0$ by employing the Lagrangian approach for the perturbation theory of soliton as Refs. \cite{Lagrangian1, Lagrangian2, Lagrangian3}, reading
 \begin{equation}
\label{Lang}
\frac{d}{dt}\left(\frac{\partial L}{\partial\dot{x}_{0}}\right)-\frac{\partial L}{\partial x_{0}}=2\text{Re}\left(\int_{-\infty}^{+\infty}R^{*}\psi\frac{\partial\psi^{*}}{\partial x_{0}}dx\right),
 \end{equation}
with $L=\int_{-\infty}^{\infty}[ \frac{i}{2}(\psi^{*}\frac{\partial}{\partial t}\psi-\psi\frac{\partial}{\partial t}\psi^{*})(1-\frac{1}{\left|\psi\right|^{2}})-\frac{1}{2}|\frac{\partial\psi}{\partial x}|^{2}-\frac{1}{2}(\left|\psi\right|^{2}-1)^{2}]dx$ being the Lagrangian and $R^{*}=(\bar{g}_{R}-\frac{i}{2}\bar{R})m_{R}$ being referred to as dissipative effects. 

The next calculations of the right-hand side of Equation (\ref{Lang}) require knowledge of the reservoir density $m_R$, In this work, we have limited 
the calculations in the fast reservoir limit, under which 
\begin{equation}
m^0_R=\frac{\bar{\gamma }_C}{\bar{\gamma}_R}\left(1-\left|\psi\right|^2\right).\label{Frate}
\end{equation} 
 
 Next,  directly following Ref. \cite{Lagrangian1}, we employ the variational method by plugging Eqs. (\ref{Tdark}) and (\ref{Frate}) into Eq. (\ref{Lang}) and obtain the equation of motion of  the relative center motion of the two-soliton as follows:
 \begin{align}
\label{EOM}
\frac{d^{2}x_{0}}{dt^{2}}&=-\frac{\partial V(x_{0})}{\partial x_{0}}+F_{\text{eff}},
\end{align}
with $V(x_{0})=2B^{4}\exp(-4x_{0}B)$ being the effective potential of  the relative center motion of the two-soliton and the dissipation-induced force $F_{\text{eff}}$, reading
\begin{eqnarray}
\label{EOM1}
F_{\text{eff}}&=&\frac{1}{3}\bar{R}\frac{\bar{\gamma}_{C}}{\bar{\gamma}_{R}}AB^{2}+\frac{4}{3}B^{3}\bar{g}_{R}\frac{\bar{\gamma}_{C}}{\bar{\gamma}_{R}}\left(8B^{2}-12Bx_{0}+9\right)e^{-4Bx_{0}}\nonumber\\
&-&\frac{32}{3}B^{7}\bar{g}_{R}\frac{\bar{\gamma}_{C}}{\bar{\gamma}_{R}}(24Bx_{0}-25)\exp(-8Bx_{0}).\label{Dforce}
\end{eqnarray}
Equation (\ref{EOM}) is the key result of this work, which allows us to interpret the dynamics of the two-dark soliton in terms of the motion of a classical particle trapped in an effective potential $V(x_{0})$ subjected to an external force $F_{\text{eff}}$.
Equation (\ref{EOM}) can recover the corresponding result in Ref. \cite{Lagrangian1} as it's expected.
From Eq. (\ref{EOM}), it is clear that the key physics is that the effective potential of $V(x_0)$ is repulsive; as a result, the two dark solitons is supposed to repel each other when they move close to each other. Based on Eq. (\ref{EOM}), we conclude that
the effective potential $V(x_0)$ which decays exponentially with the $x_0$ dominates in the case of $x_0\ll 1$, i.e., two dark solitons is relative closer to each other; in contrast, the dissipation effect captured by $F_{\text{eff}}$ plays a more important role is the case of $x_0\gg 1$, resulting in that two-dark soliton relax
by blending with the background at a finite time. 

Above, we have developed the analytically physical picture of two-dark-soliton solution based on Eq. (\ref{EOM}) and predicted features of the dissipation affecting the dynamics of two-dark soliton compared to the coherent case. In what follows,  we plan to investigate dynamics of two-dark soliton solution in a more broader parameter regimes by numerically solving equation (\ref{Spsi}) and (\ref{Srate}) with the initial wave function given by Eq. (\ref{Tdark}). In such, we focus on the interplay of nonlinearity, dispersion and dissipation affects the existence and properties of  two-dark soliton in a polariton BEC. 

We first briefly review some important features of a two-dark soliton in the coherent case, corresponding to the $m_R=0$ in Eq. (\ref{Spsi}). As a benchmark for later analysis,  Equation (\ref{EOM}) can be simplified into $d^2x_0/dt^2=-\partial V(x_0)/\partial x_0$. As a double-check of whether our analytical and numerical being correct, we compared the analytical results (solid curves) based on Eq. (\ref{EOM}) with the numerical ones (circled curves) in Figs. \ref{Fig2} (b). As it's expected that the analytical and numerical simulations are agreed with each other very well, showing that the two dark solitons repel each other because the effective potential $V(x_0)$ in Eq. (\ref{EOM}) between two dark solitons is repulsive. 

Then, we consider how the non-equilibrium nature of model affects the dynamics of two-dark-soliton solution when the dissipation parameters are turned on. In this end, we devise two scenarios: first, we choose a small initial velocity of $v_s$ and the two dark solitons will never penetrate but repel each other governed by the effective potential of $V(x_0)$ in Eq. (\ref{EOM1}) when they are closer each other; then, when the initial velocity of $v_s$ is bigger than a critical value, the two dark solitons will overcome the effectively repulsive potential and penetrate. In the first scenario, we have chosen the initial soliton's velocity with $v_s=0.15$. As shown in Fig. \ref{Fig2} (c) and (d), the relative motion's minimum value is positive due to the reduction of their repulsive force between the solitons rooted into the interaction between atoms. Moreover, the numerical results (black solid curves) based on Eq. (\ref{EOM}) find remarkable agreement with the analytically ones (blue dotted curves). Compared with Fig. \ref{Fig2} (b) without dissipation, the results with the introduction of dissipation in Fig. \ref{Fig2} (d) show that dark soltions are rebounded into farther positions, suggesting that the dissipation increases the repulsive effective potential. In the second scenario,  the initial velocity of $v_s$ is chosen to be big enough to penetrate each other as shown in Fig. \ref{Fig2} (e) and (f).  Note that our analytical results in Eq. (\ref{EOM}) are valid under the condition that the relative distance of two solitons should be bigger than  the width of soliton. Therefore our analytical results in Eq. (\ref{EOM}) are supposed to be invalid when penetrating each other. In contrast,  before and after collision of two dark solitons correspond to the relative distance of two solitons being bigger than  the width of soliton, our analytical results in Eq. (\ref{EOM}) are found to be consistent with the numerically ones as shown in Fig. \ref{Fig2} (e) and (f).

{\it Conclusion.---} In summary, we have investigated the dynamics of two dark solitons appearing in polariton BECs under nonresonant pumping. We have derived analytically the evolution equations for the solitons parameters. Within the framework of Lagrangian approach, our analytical results capture the essential physics about how the open-dissipative the effects affects the relative motion of two solitons at a finite time. We also solve the dissipative equation by the initial wave function of two dark solitons in a numerically exact fashion. The numerical results find remarkable agreement with the analytically ones. We also have investigated the collision of two solitons in polariton BECs under nonresonant pumping. By manipulating the initial velocity, the relative motion of the two solitons can repel or penetrate each other.

We thank Alexey Kavokin and Ying Hu for stimulating discussions. This work was
supported by Zhejiang Provincial Natural Science Foundation of China (Grant No. LZ21A040001), the National Natural
Science Foundation of China (No. 12074344) and the key projects of the Natural Science Foundation of China (Grant No. 11835011).
\bibliography{apssamp}
\end{document}